\documentclass[journal=jacsat,manuscript=article]{achemso}
\usepackage{palatino}
\usepackage[latin1]{inputenc}
\usepackage{epsf}
\usepackage{amsmath,amssymb}
\usepackage{latexsym}
\usepackage{calc}
\usepackage{color}
\usepackage{graphicx}

\def\sss{\scriptscriptstyle\rm}

\def\subext{_{\sss ext}}

\def\subc{_{\sss C}}
\def\subh{_{\sss H}}
\def\subs{_{\sss S}}
\def\subxc{_{\sss XC}}
\def\supS{^{\rm S}}

\def\1s{_{1,\sss S}}
\def\2s{_{2,\sss S}}

\def\c{_{\sss C}}
\def\s{_{\sss S}}
\def\xc{_{\sss XC}}

\def\H{_{\sss H}}

\def\ext{_{\rm ext}}
\def\S{^{\rm S}}

\def\cT{{\cal T }}

\def\trho{\tilde{\rho}}
\def\tk{\tilde{k}}
\def\tn{\tilde{n}}

\def\br{{\bf r}}
\def\bj{{\bf j}}
\def\bb{{\bf b}}

\author{Lionel Lacombe}
\affiliation{Department of Physics and Astronomy, Hunter College and the Graduate Center of the City University of New York, 695 Park Avenue, New York, New York 10065, USA}
\email{liolacombe@gmail.com}
\author{Neepa T. Maitra}
\affiliation{Department of Physics and Astronomy, Hunter College and the Graduate Center of the City University of New York, 695 Park Avenue, New York, New York 10065, USA}
\email{nmaitra@hunter.cuny.edu}

\title{Density-matrix coupled time-dependent exchange-correlation functional approximations}

\abbreviations{}

\date{\today}

\begin{document}

\begin{abstract}
We present a new class of non-adiabatic approximations in time-dependent density functional theory derived from an exact expression for the time-dependent exchange-correlation potential. The approximations reproduce dynamical step and peak features in the exact potential that are missing in adiabatic approximations. Central to this approach is an approximation for the one-body reduced density-matrix as a functional of the Kohn-Sham density-matrix, and we demonstrate three such examples. 
\end{abstract}

\maketitle

\section{Introduction}

Time-Dependent Density Functional Theory (TDDFT) is in principle an exact reformulation of many-electron quantum mechanics~\cite{RG84}, where non-interacting electrons evolve in a one-body potential, the Kohn-Sham (KS) potential, such that the exact one-body density $n(\br,t)$ of the true system is reproduced. In practice, the exchange-correlation (xc) contribution to the potential must be approximated, as a functional of the density, the initial interacting state $\Psi(0)$ and the initial KS state $\Phi(0)$: $v\xc[n;\Psi(0),\Phi(0)](\br,t)$. Almost all TDDFT calculations today use an adiabatic approximation, in which the instantaneous density is simply inserted into a chosen ground-state functional approximation: $v\xc^{\rm A}[n; \Psi(0),\Phi(0)](t) = v\xc^{\rm g.s.}[n(t)](t)$. With such an approximation, TDDFT has become a widely-used and trusted method for electronic spectra and response~\cite{M16,AJ13}, and users are generally aware to be cautious for cases  (such as double-excitations) where the adiabatic approximation is known to fail. 

Over the past decade, TDDFT applications in the non-perturbative regime have grown in depth and breadth, from attosecond charge migration~\cite{BHMALGSL18}, to photovoltaic design~\cite{RFSR13}, to ultrafast demagnetization~\cite{EMDSG16}, for examples. Many of these calculations yield results in reasonable agreement with experiment, and allow interpretation of mechanisms involved in the various processes. 
There are, on the other hand, cases where the TDDFT calculations perform poorly, as in Refs.~\cite{RN11,RN12b,RN12c,HTPI14,WU08,GDRS17} and likely more such cases go unpublished, and recent work has strongly suggested this is often due to the errors inherent in the ubiquitous adiabatic approximation for the xc functional~\cite{EFRM12,RG12,FERM13,SLWM17,M17,LSWM18,CPRS18,M16,FLSM15}. Indeed, it is surprising that the adiabatic approximation has been as successful as it has been when the system wavefunction evolves far from a ground-state; this could possibly be due to its trivial satisfaction of exact conditions involving memory, while attempts to introduce memory can break these exact conditions~\cite{GK85,D94b,V95,Vignalechap,M16}, or are based on the response of the (weakly inhomogeneous) uniform gas that results in endowing finite systems with artificial linewidths and damping~\cite{VK96,VUC97,UB04}. 

In this work, we introduce a class of approximations derived from an exact expression for the xc potential. The approximations do not rely on any variational principle, immediately break free of the adiabatic approximation, and can be viewed as functionals of the xc hole and one-body reduced density-matrix (1RDM) of the KS system. They have a form that produces the elusive dynamical step and peak features of the correlation potential missing in the adiabatic approximations, even for simple approximations used for the 1RDM.

\section{Density-matrix coupled approximations based on the exact exchange-correlation potential}

The exact expression for the xc potential is derived from equating the equation of motion for the density of the interacting system,
\begin{multline}
  \frac{\partial^2 n(\br,t)}{\partial t^{2}} 
  =
  \nabla\cdot \left(
    \frac{1}{4}(\nabla'-\nabla)(\nabla^2-\nabla'^2)\rho_1(\br,\br',t)\vert_{\br=\br'} \right.\\
    + \left.\nabla v\subext(\br,t)n(\br,t) 
    + \int d^3\bar{\br}\nabla w(\br,\bar{\br}) \rho_{2}(\br,\bar{\br};\br,\bar{\br},t)
  \right) 
  \label{Eq:theo_d2n_corr}
\end{multline}
to that of the non-interacting KS system~\cite{L99,LFSEM14,FLNM18}. Here the 1RDM 
\begin{equation}
  \rho_1(\br,\br') 
  =
  N\int dr_{2}..dr_{N}
  \Psi^*(\br',\br_{2},..,\br_{N})\Psi(\br ,\br_{2},..,\br_{N})
\end{equation}
, and the two-body reduced density-matrix (2RDM)  
\begin{equation}
\rho_2(\br_{1},\br_{2},\br_{1}',\br_{2}') 
=
N(N-1)\int dr_{3}..dr_{N}
\Psi^*(\br_{1}',\br_{2}',\br_{3},..,\br_{N})
\Psi(\br_{1},\br_{2},\br_{3},..,\br_{N})
\quad.
\end{equation}
Using the definition $v\s(\br,t) = v\ext(\br,t) + v\H(\br,t) + v\xc(\br,t)$ 
this yields an expression for the exchange-correlation potential:
\begin{equation}
\nabla\cdot\left(n(\br,t)\nabla v\subxc(\br,t)\right) 
    =\nabla\cdot(n(\br,t)\nabla v\subxc^W(\br,t))  
    +   \nabla\cdot(n(\br,t)\nabla v\subc^T(\br,t))
\end{equation}
where the interaction component $v\xc^W$ satisfies
\begin{equation}
  \nabla\cdot(n(\br,t)\nabla v\subxc^W(\br,t)) =  
  \nabla\cdot\Big[
    n(\br,t)\int n\subxc(\br',\br,t)
    \nabla w(\vert\br'-\br\vert)d^3r'
  \Big]\,,
  \label{Eq:theo_vxcw}
\end{equation}
and the kinetic component $v\c^T$ satisfies:
\begin{equation}
  \nabla\cdot(n(\br,t)\nabla v\subc^T(\br,t)) =  
  \nabla\cdot\Big[\mathcal{D}
    (\rho_1(\br',\br,t)-\rho\1s(\br',\br,t))|_{\br'=\br} 
  \Big] \,,
  \label{Eq:theo_vct}
\end{equation}
with $\mathcal{D} = \frac{1}{4}(\nabla'-\nabla)(\nabla^2-\nabla'^2)$. 
Here $n\subxc$ is the time-dependent xc hole defined as
\begin{equation}
  n\subxc(\br,\br') = \rho_{2}(\br,\br';\br,\br')/n(\br') - n(\br)
  \quad,
  \label{Eq:theo_nxc}
\end{equation}
and $\rho\1s$ is the 1RDM of the KS system.

To be of any practical use, approximations must be made in Eqs.~(\ref{Eq:theo_vxcw})--(\ref{Eq:theo_vct}) for the terms  involving the correlated interacting wavefunction, i.e. $n\xc$ and $\rho_1$, as functionals of the initial interacting state and quantities accessible in a KS evolution; such quantities are implicit functionals of the density, including its history, and the KS initial state. 

A first approximation, which has been previously explored~\cite{RB09b,FNRM16,SLWM17,LSWM18,FLNM18,LHRC17}, is to replace both $\rho_1$ and
$n\xc$ by their KS counterparts. In this approximation, which we dub $v\xc\S$, the approximate interaction component is generally a very good approximation to the exact $v\xc^W$, but the kinetic component vanishes. Although in general it is non-adiabatic, the dynamical step and peak features are absent, since these appear in $v\c^T$. These features have been shown to be crucial to accurately capture dynamics in a number of situations, e.g. electron scattering~\cite{SLWM17,LSWM18}, charge-transfer out of a ground state~\cite{FERM13}, quasiparticle propagation through a wire~\cite{RG12}. An approximation for $v\c^T$ is essential. 


To this end, we first introduce a 
fictitious system
represented by a 1RDM $\trho_1(t)$, with $\trho_1(0) = \rho_1(0)$ provided by the initial interacting wavefunction,  which follows the equation of motion
\begin{equation}
    i\frac{\partial \trho_{1}(\br,\br',t)}{\partial t} 
    = 
    [-\nabla^2/2 + v\ext,\trho_{1}] (\br,\br',t)
    + \int d^3\bar{\br} 
    (w(\br,\bar{\br})-w(\br',\bar{\br})) 
    \trho_{2}[\trho_{1},\rho\1s] (\br,\bar{\br};\br',\bar{\br})
    \label{eq:rho1tilde}
\end{equation}
where we consider $\trho_{2}$ as a functional of $\trho_{1}$ and $\rho\1s$. This equation has the form of the first equation in the so-called BBGKY heirarchy for density-matrices~\cite{Bonitzbook}. 
The problem then becomes one of finding an approximation to $\trho_2[\trho_{1},\rho\1s]$. Armed with such an approximation, one can proceed in two equivalent ways.  In one, the approximate $\trho_2[\trho_{1},\rho\1s]$ is used at each time-step in Eq.~(\ref{eq:rho1tilde}) to determine $\trho_1(t)$ through propagation for input into Eq.~(\ref{Eq:theo_vct}) for $v\c^T$ and also into the right-hand-side of Eq.~(\ref{Eq:theo_nxc}) for input into Eq.~(\ref{Eq:theo_vxcw}) for $v\xc^W$.
In terms of the explicit dependences, one can write:
\begin{align}
  v\subxc^W[n(t),n\subxc(t)] 
  &\rightarrow 
  v\subxc^W[\tn(t),\tn\subxc(t)] \\
  v\subc^T[n(t),\rho_{1}(t)-\rho\1s(t)] 
  &\rightarrow 
  v\subc^T[\tn,\trho_{1}(t)-\rho\1s(t)]
  \quad
\end{align}
and, with the potentials defined in this way, $\trho_1(\br,\br,t) = \rho\1s(\br,\br,t) = \tn(\br,t)$ (see shortly). 
Here $\tn\subxc(t)$ is defined as in Eq.~(\ref{Eq:theo_nxc}) but with $\trho_{2}[\trho_{1},\rho\1s]$.
One propagates Eq.~(\ref{eq:rho1tilde}) alongside the TDKS equation,
\begin{equation}
i \partial_t\phi_i(\br,t) = \left(-\nabla^2/2 + v\ext(\br,t) + v\H(\br,t) + v\xc(\br,t)  \right)\phi_i(\br,t)\,,
\label{eq:tdks}
\end{equation}
with the resulting $\rho\1s$ input into $\trho_2[\trho_1,\rho\1s]$, which in turn determines $\tn\xc$ and  $\trho_1$ through Eq.~(\ref{eq:rho1tilde})
to obtain the KS potentials for Eq.~(\ref{eq:tdks}).

From an alternative but equivalent view, the diagonal of the $\trho_1$ obtained from propagation of Eq.~(\ref{eq:rho1tilde}) with the approximate $\trho_2[\trho_{1},\rho\1s]$ yields a density, $\tn(\br,t)$, which is  required to be the density of the KS system. That is, $v\xc^W$ and $v\c^T$ are defined such that, once added to $v\ext$ and the Hartree potential $v\H$, result in a non-interacting KS system that has density $\tn(\br,t)$. This requires, alongside Eq.~(\ref{eq:rho1tilde}), an inverse problem of a similar nature to that considered in Ref.~\cite{NRL13}, to be solved at each time-step to obtain the potentials to propagate the TDKS system, from which $\rho\1s$ is extracted. 

That these two ways give the same result, can be seen by applying Eq.~(\ref{Eq:theo_d2n_corr}) to the fictitious system,  where $n \rightarrow \tn$, 
$\rho_{1} \rightarrow \trho_{1}$ and 
$\rho_{2} \rightarrow \trho_{2}[\trho_{1},\rho_{1,s}]$ are used.
Forcing $\partial^2_t n\s$ and $\partial^2_t \tn$ to be equal results in
equations~(\ref{Eq:theo_vxcw}) and~(\ref{Eq:theo_vct}) 
for $v\subxc^W$ and $v\subc^T$ but in terms of $\tn$, $\tn\subxc[\tn,\trho_2]$ 
and $\trho_1$.


A crucial aspect in the definition of $\trho_{2}[\trho_{1},\rho\1s]$ is the
dependence on $\rho\1s$. 
If $\trho_{2}$ is taken as a functional of only $\trho_{1}$ then the propagation 
of $\trho_1$ is simply that of reduced-density matrix functional theory (RDMFT), and the KS system would be simply the non-interacting system that reproduces
this RDMFT density, if it exists.
In this situation, the KS propagation is redundant, as all the information comes from the RDMFT. Approximations in time-dependent RDMFT still struggle with much of the same problems that adiabatic approximations in TDDFT struggle with~\cite{EM16}. 
The point of this paper is  instead to focus on TDDFT with 
KS propagation, where $\trho_1$ is used only to provide an approximation for $v\subc^T$. 
Contrary to RDMFT,  $\trho_1$ does not need to be $N$-representable as it is not intended 
to be a physical quantity, but instead its purpose is as a calculation tool for our KS system. The RDMs $\trho_2$ and $\trho_1$ are not contraction-consistent, and only the near-diagonal elements of $\trho_1$ have some physical meaning. 

For a first approach within this class of approximations we take 
\begin{equation}
\trho_2[\trho_1,\rho\1s] \to \rho\2s\,,
\label{eq:rho2s_app}
\end{equation}
the 2RDM of the 
 KS system. 
 In fact, the interaction component then reduces to the $v\xc\S$ approximation: 
$  v\subxc^W[n(t),n\subxc(t)] \to 
  v\subxc^W[n(t),n\S\subxc(t)] 
  \equiv
  v\xc\S$, but, importantly, the $v\c^T$-component is no longer zero. We denote the xc potential arising from this approximation as $v\xc^{\trho}$. 

We also briefly discuss a second approximation, denoted HF-driven, which is in a sense complementary to that of Eq.~(\ref{eq:rho2s_app}): 
instead of having no explicit dependence of $\trho_2$ on the fictitious 1RDM $\trho_1$, HF-driven ignores the $\rho\1s$ dependence and takes $\trho_2$ as the Hartree-Fock (HF) functional of $\trho_1$. That is
\begin{equation}
\trho_2[\trho_1,\rho\1s] \to \rho_2^{\rm HF}[\trho_1]\,,
\label{eq:HF-driven}
\end{equation}
which, for a spin-singlet, gives $\trho_2(\br_1,\br_2;\br_1',\br_2') =
\trho(\br_1,\br_2)\trho(\br_1',\br_2')-\frac{1}{2}\trho(\br_1,\br_2')\trho(\br_1',\br_2)$.
The KS potential found is that which reproduces the density of the TDHF evolution of an initially-correlated wavefunction. As discussed above, such an approximation would not yield a useful method for TDDFT, and its only purpose here is to illustrate that dynamical step and peak features in $v\xc$ can be retrieved with relatively simple approximations for $\trho_2[\trho_1,\rho\1s]$.

Before presenting the numerical results on some model systems, we point out that $v\xc^{\trho}$ satisfies two fundamental exact conditions in TDDFT. 
 Perhaps most important is the Zero Force Theorem (ZFT)~\cite{GDP96,V95b,V95,D94b,Vignalechap}, since violation of this can lead to self-excitation and numerical instabilities~\cite{MKLR07}.  The ZFT states that the xc potential cannot exert a net force, $\int n(\br,t)\nabla v\xc(\br,t) d^3 r =0$, and in Ref.~\cite{FLNM18} it was shown that $v\xc\S$ satisfies this. It remains to investigate whether $v\c^{T,\trho}$ (i.e. $v\c^T$ of Eq.~(\ref{Eq:theo_vct}), with $\rho_1$ computed as in Eq.~(\ref{eq:rho1tilde})  with Eq.~(\ref{eq:rho2s_app})), also satisfies the ZFT.
First, we write $v\subc^{T,\trho}$ as the difference of two terms $v_{{\sss C},int}^{T,\trho}$
and $v_{{\sss C},S}^{T,\trho}$ defined as:
\begin{equation}
\nonumber
\nabla\cdot(n\nabla v_{{\sss C},int}^{T,\trho}) = \nabla\cdot
    \Big[\mathcal{D}
      \trho_1(\br',\br,t)|_{\br'=\br} 
    \Big] 
\end{equation}
with 
$v_{{\sss C},S}^{T,\trho}$ defined in the same way but with $\trho\1s(\br',\br,t)$ on the right-hand-side.
In fact, $v_{{\sss C},int}^{T,\trho}$ and $v_{{\sss C},S}^{T,\trho}$ independently satisfy the ZFT; we show the proof for $v_{{\sss C},int}^{T,\trho}$ only as that for $v_{{\sss C},S}^{T,\trho}$ follows analogously.
Writing $\trho_{1}(\br,\br')$ as
\begin{equation}
  \trho_{1}(\br,\br') = \sum_i \tilde \omega_i \tilde \phi_i^*(\br') \tilde \phi_i(\br)\,,
\end{equation}
we compute 
\begin{eqnarray}
\nonumber
\int  n(\br,t){\bf \nabla} v_{{\sss C},int}^{T,\trho} (\br,t)d^3r = 
  \int d^3r \mathcal{D}\trho_{1}(\br',\br,t)\vert_{\br'=\br} 
    \\
    = \frac{1}{4} \sum_i \tilde \omega_i \int d^3r 
    \nabla \left(4\vert \nabla\tilde\phi_i\vert^2 - \nabla^2\vert\tilde\phi_i\vert^2\right)\,.
    \end{eqnarray}
which involves the values of the orbitals and derivatives at the boundary. 
The integral thus vanishes for finite systems, or if the system is periodic. Hence, $v\subc^{T,\trho}$ satisfies the zero-force theorem.

Another important theorem in TDDFT is Generalized Translation Invariance (GTI), which states that in a uniformly boosted frame where 
$  |\Psi^\bb (\br_1...\br_N,t) \rangle
  =  \prod^N_{j=1}e^{-i \br_j \cdot\dot{\bb}(t)}
  |\Psi (\br_1+\bb(t)...\br_N+\bb(t),t)\rangle
$, with $\bb(0) = \dot\bb(0)= 0$, the xc potential transforms as
\begin{equation}
  v\subxc^{\bb}[n; \Psi(0), \Phi(0)](\br,t) = v\subxc[n;\Psi(0), \Phi(0)](\br+\bb(t),t)
  \quad.
\end{equation}
Ref.~\cite{FLNM18} proved that $v\xc\S$ satisfies the GTI, so it again remains to show that $v\c^{T,\trho}$ does. That it does follows along the same proof as that given in Ref.~\cite{FLNM18}, where 
\begin{equation}
\nabla\cdot\left(n^\bb(\br(t),t) \nabla v\subc^{T,\trho,\bb}(\br,t)\right)  = \nabla\cdot\left(n(\br^\bb(t),t) \nabla v\subc^{T,\trho}(\br +\bb(t),t)\right)
\end{equation}
The proofs for the HF-driven TDDFT of Eq.~(\ref{eq:HF-driven}) follow in a similar way. 

We also note that $v\xc^{\trho}$ is one-electron self-interaction free:
$v\subxc\S$ cancels $v\subh$ and $v\subc^{T,\trho}$ is 
zero since $\rho_1 = \rho_{1,s} = \trho_1$ for one electron. 

\section{Numerics}


Turning now to the numerical implementation of the approach, one requires two propagations: one is Eq.~(\ref{eq:rho1tilde}) to obtain $\trho_1$, and the other is the KS equation Eq.~(\ref{eq:tdks}). These need to be run self-consistently, with $\trho_1$ obtained from the KS orbitals of Eq.~(\ref{eq:rho1tilde}) feeding into the potentials in Eq.~(\ref{eq:tdks}) and $\rho\2s$ from Eq.~(\ref{eq:tdks}) feeding into Eq.~(\ref{eq:rho1tilde}).
We have found that because of the multiple derivatives involved  and the sensitivity to error in the potential, using
the formula in Eq.~(\ref{Eq:theo_vct}) directly is numerically challenging.
Even using the exact time-dependent quantities, a straightforward implementation 
of this expression does not generate the exact density. Instead, we follow the equivalent approach, of requiring that 
the KS potential $v\s$ is such that Eq.~(\ref{eq:tdks}) reproduces the density $\tn$ at each time-step.
Thus we invert numerically
the equation of motion of $\rho\1s$ to find the corresponding potential.
In the following we present the main concepts we used to propagate the two quantum systems 
while forcing their densities to stay equal. 
We restrict ourselves here in the discussion to a simple forward Euler scheme to present the ideas that can be
generalize to any multi-step schemes.

\subsection{KS system}
The equation of evolution for the density matrix $\rho\1s$ reads
\begin{equation}
  \rho\1s^{+1} = \rho\1s - i[T,\rho\1s]dt - i[v,\rho\1s]dt
  \label{eq:discretedrhodt}
\end{equation}
where $\rho\1s^{+1}$ is the density matrix at the next time-step.
If one only considers the diagonal terms it yields :
\begin{equation}
  n\subs^{+1} = n\subs - i[T,\rho\1s]\vert dt 
  \label{Eq:num_n_prop}
\end{equation}
where $A\vert$ stands for $A\vert_{\br=\br'}$ and $n=\rho\1s\vert$.
We also introduce two notations.
The first one is
\begin{equation}
  -i[T,\rho\1s] \vert = \cT \rho\1s
  \label{Eq:num_F_def}
\end{equation}
where $\cT$ is a linear operator of $\rho\1s$ (hence the product notation).
$\cT \rho\1s$ can be identified as $-\nabla\cdot\bj$.
One can note that $n^{+1}$ is independent of $v$ and can be obtained
without computing it.
The second notation we introduce is $k=\partial \rho\1s/ \partial t$ so that Eq.~(\ref{eq:discretedrhodt}) can be compactly written as
\begin{equation}
  \rho\1s^{+1} = \rho\1s + k\,dt 
  \quad.
\end{equation}
Of course it is easy to verify $k\vert = \cT \rho\1s$.
The matrix $k$ has a linear dependence on the potential $v$ at time $t$ : $k=k(v)$.

\subsection{Reference system}
For the reference system $\trho_1$ we start with the first equation of the BBGKY hierarchy:
\begin{equation}
  \trho_1^{+1} = \trho_1 - i[T,\trho_1]dt - i[v,\trho_1]dt - i\int(w-w)\trho_2 dt
\end{equation}
where $\int(w-w)\trho_2$ is a short notation for
$\int d\bar{\br}(w(\br,\bar{\br})-w(\br',\bar{\br}))\trho_2(\br,\bar{\br},\br',\bar{\br})$.
More compactly, we will denote
\begin{equation}
  \trho_1^{+1} = \trho_1 + \tk\,dt
\end{equation}
while
\begin{equation}
  \tn^{+1} = \tn + \cT \trho_1 dt 
  \label{Eq:num_tn_prop}
  \quad.
\end{equation}

\subsection{Time propagation}
If at time $t$ one assumes
\begin{align}
  n\subs &= \tn \\
  \cT\rho\1s &= \cT\trho_1
\end{align}
then $n^{+1}=\tn^{+1}$. 
$\cT k$ can be identified as $-\nabla\cdot\frac{\partial\bj}{\partial t}$.
If $\cT k(v) = \cT \tk$ we obtain $\cT \rho\1s^{+1} = \cT \trho_1^{+1}$ which ensures that 
$n^{+2} = \tn^{+2}$. 
So we  choose $v$ so that $\cT k(v) = \cT \tk$ which amounts 
to solving a linear system of equations for $v$ as both $\cT$ and $k$ are linear mappings.
To do so, we minimize 
\begin{equation}
  || \cT k(v) - \cT \tk ||^2 + \varepsilon ||\nabla v||^2
  \label{Eq:num_min}
\end{equation}
for $v$, where $\varepsilon$ is a small regularization parameter that makes us approach the solution
with the smoothest potential possible. Note that $\nabla v$ can be replaced by $\nabla^2 v$ with no significant difference in the result.
This procedure yields an error in the density proportional to $\varepsilon$. 
In the following we chose to use this technique with
an explicit fourth-order Runge-Kutta method.


\section{Results}

To test the density-matrix coupled approximation Eq.~(\ref{eq:rho2s_app}), we studied a model system of two soft-Coulomb interacting electrons in 1D.
Such a simple model allows us to have access to the exact solution and the exact KS
potential, while retaining essential features of the exact 3D Coulomb-interacting problem~\cite{VIC96,BS02,LGE00,GW14}. 
The external potential is $v\subext(x) = -2/\sqrt{1+x^2}$ while the interaction is  $w(x,x') = 1/\sqrt{1+(x-x')^2}$. 
In the following, $\Psi_0$ and $\Psi_1$ refer to the ground and first excited singlet
states of the interacting Hamiltonian respectively.

First we consider field-free evolution of the  state
\begin{equation}
  \Psi(t=0) = (\Psi_0 + \Psi_1)/\sqrt{2} 
  \quad,
  \label{Eq:res_50:50}
\end{equation}
resulting in a density that is periodic in time, with a period of $2\pi/(E_1 - E_0) = 11.788$ a.u.
This initial wavefunction defines $\trho_1(0)$ for the initial condition  of Eq.~(\ref{eq:rho1tilde}). 
The TDDFT KS system may begin in any initial state that
reproduces the density $n(0)$ of Eq.~(\ref{Eq:res_50:50}) and its time-derivative $\partial_t n(0)$.
We consider choices of the form
\begin{equation}
  \Phi(t=0) = \frac{1}{\sqrt{1+a^2}}(\Phi_0^{(a)} + a\Phi_1^{(a)}) 
\end{equation}
with 
\begin{eqnarray}
\nonumber
  \Phi_0^{(a)}(x_1,x_2) &= &\phi_0^{(a)}(x_1)\phi_0^{(a)}(x_2)
  \\
  \nonumber
  \Phi_1^{(a)}(x_1,x_2) &= &
  \frac{1}{\sqrt{2}}(\phi_0^{(a)}(x_1)\phi_1^{(a)}(x_2) +
\phi_1^{(a)}(x_1)\phi_0^{(a)}(x_2))
\nonumber
\end{eqnarray}
The single-particle orbitals $\phi_i^{(a)}$ are computed so that $\Phi(t=0)$ has
the right initial conditions for the density.
Here, we will discuss the cases $a=1$ and $a=0$.

\begin{figure}[htbp]
  \centering
  \includegraphics[width=1.0\columnwidth]{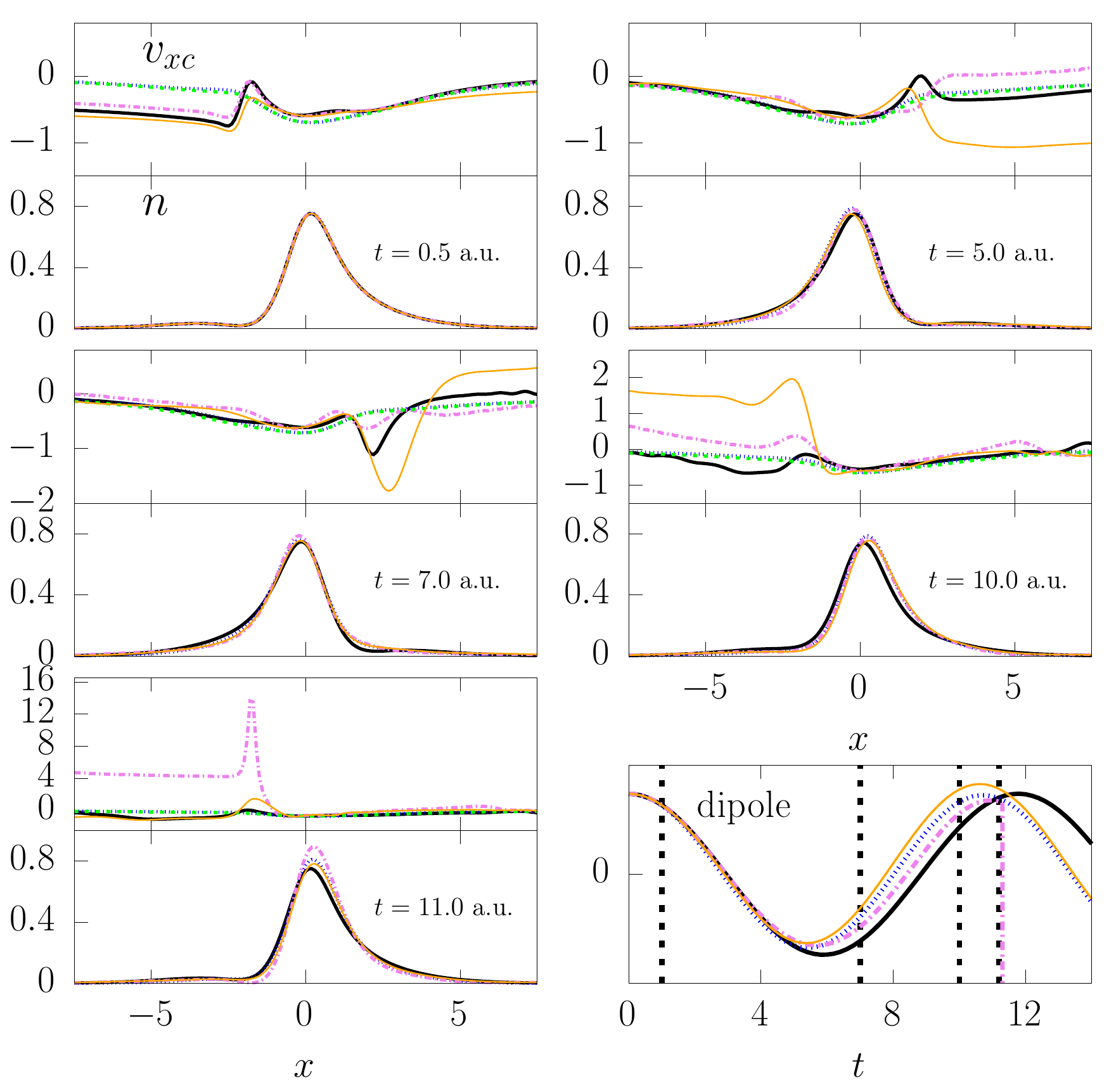}
  \caption{
    Time-snapshots of the xc potential (upper subpanels) and density (lower panels) for field-free 
     propagation of the state Eq.~(\ref{Eq:res_50:50}), predicted by KS evolution of $a=1$ KS state: exact $v\subxc$ (black solid), exact $v\subxc^W$ (green dashed),
    $v\subxc\S$ approximation (blue dotted), $v\subxc^{\trho}$ (pink dot-dashed), and HF-driven (thin orange solid). 
     The lowest right panel shows the corresponding dipole moments, with
   vertical dashed lines indicating times shown in the other  panels.
  }
  \label{Fig:ffa1pot}
\end{figure}

We first consider the case $a=1$. 
Figure~\ref{Fig:ffa1pot} shows the potential and density obtained by the approximations 
$v\subxc\supS$  and $v\subxc^{\trho}$ 
alongside the exact, and, for illustration, the HF-driven; a movie is provided in the supplementary information. The figure also shows the $v\subxc^W$ component of the exact potential. 

Initially, since $\trho_1(0)=\rho_1(0)$, the $v\subc^{T,\trho}(0)$ component of the potential $v\subxc^{\trho}(0)$ is 
equal to  the exact $v\subc^T(0)$. The density oscillates from side to side, with its small  bump at $x\approx\pm 4$ a.u.
As it evolves, the potential $v\subxc^{\trho}$ reproduces 
qualitatively the peaks and steps of the exact potential, although not exactly in phase.
On the other hand, $v\subxc\supS$ stays relatively smooth; as
in previous work~\cite{LFSEM14,FNRM16,SLWM17} the
complex structures of the exact potential come from $v\subc^T$, while $v\subxc\supS$ cradles the density and 
closely mimics $v\subxc^W$ at all times. The steps and peaks are also reproduced by the simple HF-driven propagation, showing that they arise with even the simplest non-zero approximation for $\trho_1 - \rho\1s$; crucial is the structure of $v\c^T$ itself. 
However, after some time in the propagation, $v\subxc^{\trho}$ develops a fatal problem:
self-amplifying needle-like peak structures appear that go to infinity causing the propagation to collapse. An example of this structure is visible at $t=11$ a.u. just before the calculation collapses. As we will shortly argue, the development of such non-$v$-representable structures in the  $v\subxc^{\trho}$ approximation is not a numerical feature, but rather is intrinsic to the approximation itself; it does not show up in the HF-driven potential. 
The effect of the  peaks and steps on the density is difficult to discern in Fig.~\ref{Fig:ffa1pot}, but the dipole shown in the lowest right panel
shows the improvement of $v\subxc^{\trho}$   over $v\subxc\supS$ before the collapse. 


We also studied the case $a=0$ which is more challenging for TDDFT, as the initial KS state
is a ground-state with 
a structure totally different from the interacting state.
The initial step in the exact potential is larger than for $a=1$ as evident in Fig.~\ref{Fig:ffa0} and in the movie in supplementary material. 
The kinetic component $v\subc^T$ is crucial to obtain correct dynamics
in this situation, and propagation with $v\subxc\supS$ does not provide a good
dipole moment (lower right panel), although it is  still close to the exact $v\subxc^W$.
Our approximation $v\xc^{\trho}$ gives a better dipole and density, approximately capturing step structures, but
 again, after some time, develops the sharp needle-like structure that goes to infinity, killing the propagation. Again the HF-driven potential is shown to illustrate that this simplest approximation for $\trho_1 - \rho\1s$ also reproduces step structures (and its propagation remains stable). 
 
\begin{figure}[htbp]
  \centering
  \includegraphics[width=1.0\columnwidth, height = 0.7\columnwidth]{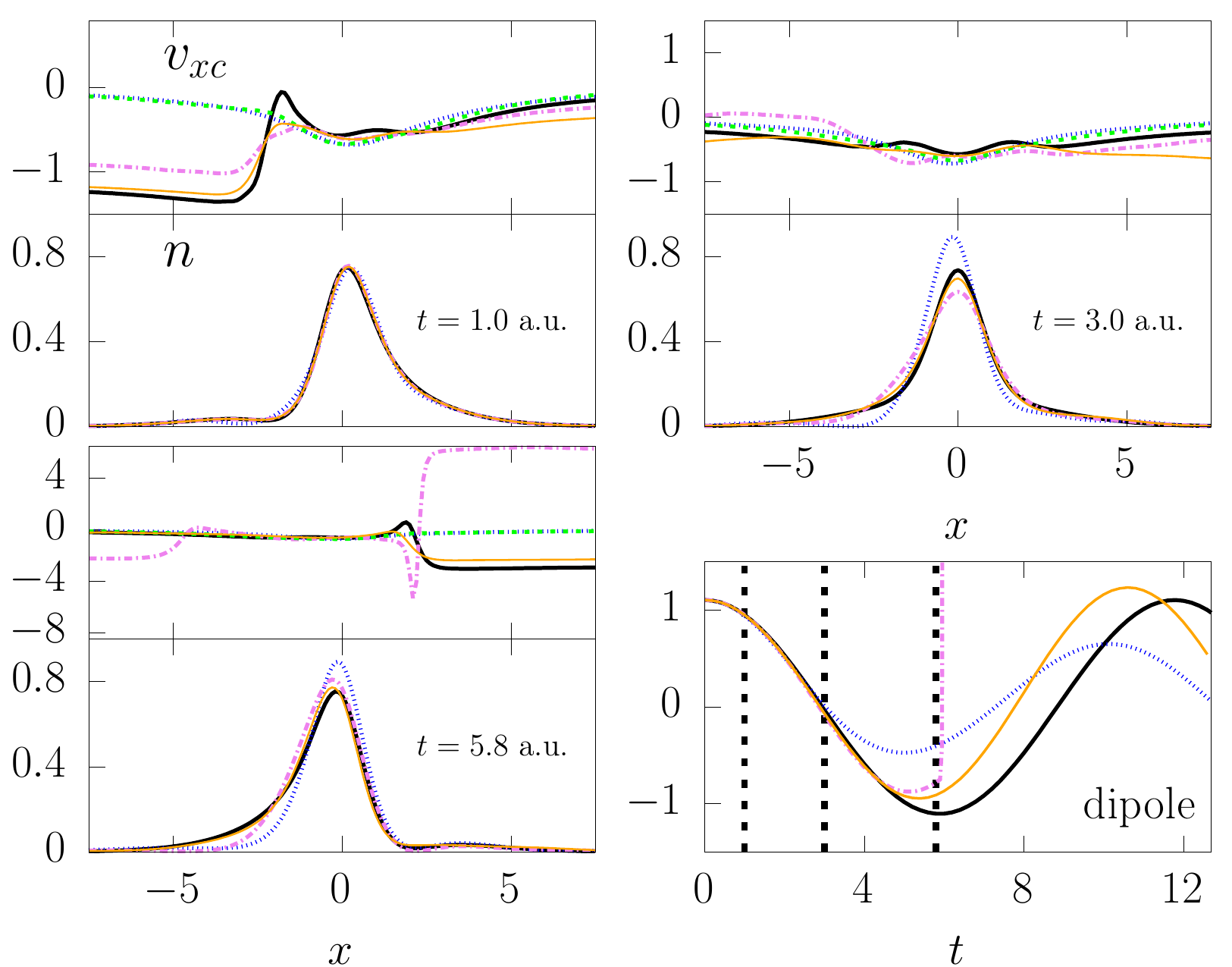}
  \caption{ As in Fig.~\ref{Fig:ffa0} but with KS evolution of the $a=0$ KS state.
  }
  \label{Fig:ffa0}
\end{figure}

\begin{figure}[htbp]
  \centering
  \includegraphics[width=1.0\columnwidth]{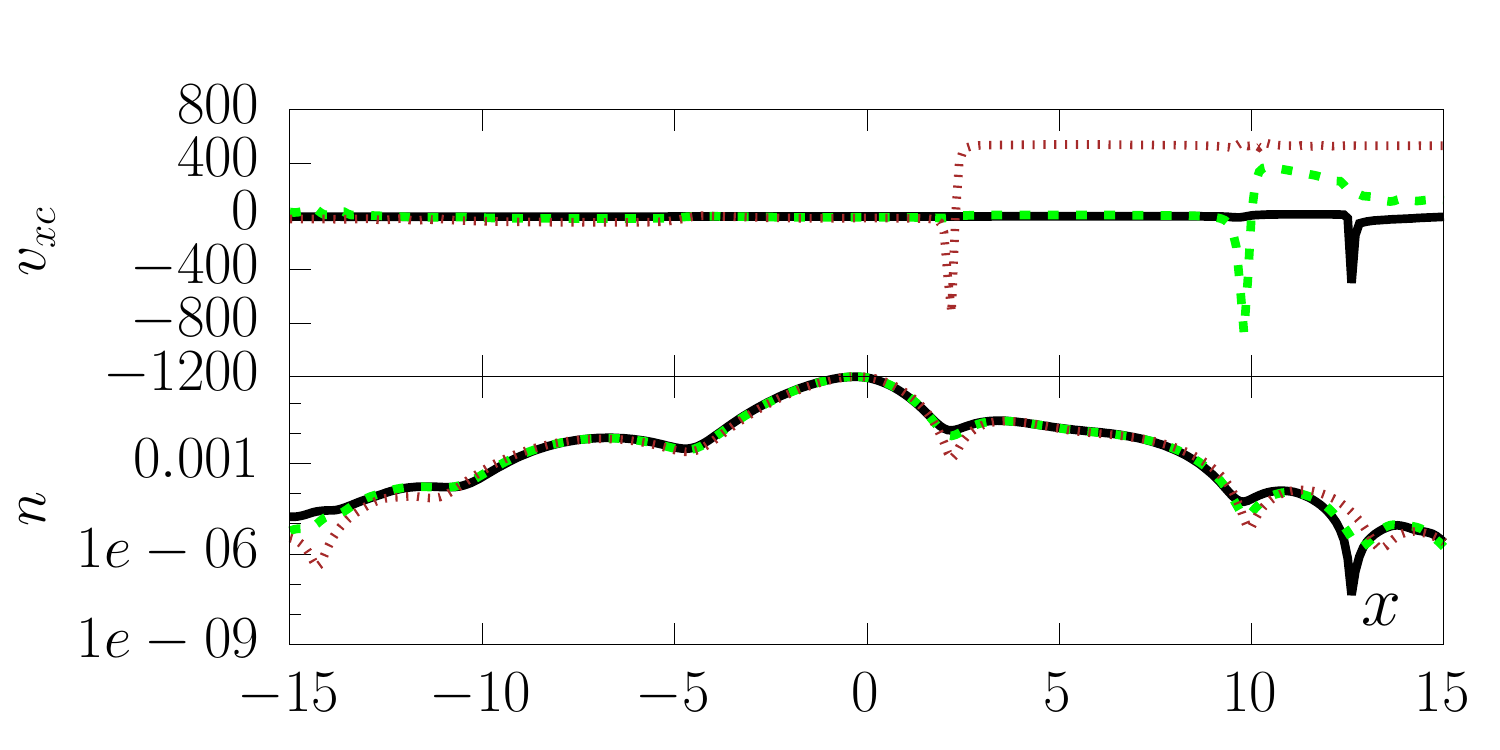}
  \caption{
    The potential $v\subxc^{\trho}$ (upper panel) and logarithm of the density (lower panel) of the calculation of Fig.~\ref{Fig:ffa0}, at its final time $t_f$,  with different smoothing parameters: $\varepsilon = 10^{-6}$ (black solid, $t_f = 5.45$ a.u.),
     $\varepsilon = 10^{-5}$ (green dashed, $t_f = 5.6$ a.u, with $v\subxc^{\trho}$ scaled by a factor of 10), 
     $\varepsilon = 10^{-4}$ (brown dotted, $t_f = 5.9$ a.u,  with $v\subxc^{\trho}$ scaled by a factor of 20).
  }
  \label{Fig:ffa0_epsilon}
\end{figure}

To see if we can numerically get rid of these needles we varied
the smoothing parameter $\varepsilon$. 
In Fig.~\ref{Fig:ffa0_epsilon} we plot the density in logscale 
(lower panel) and the potential
$v\subxc^{\trho}$ (upper subpanels) just before our calculation collapses.
For the smaller $\varepsilon$, the potential creates a needle in the region of very low
density, where the logarithmic density-derivative is almost discontinuous. 
One can assert that low density region is not so important and increase $\varepsilon$ but
this just enables the propagation to go a little further until it creates another needle at
the next derivative "discontinuity". 
Even with  bigger $\varepsilon$ we cannot get rid of 
this infinite peak that now appears well inside the region of interest. 
Increasing $\varepsilon$ further would have an impact
on the large density regions and drastically alter the intention of the approximation.
This behavior and the fact that reducing $dt$ does not improve our calculation
suggest that the needles are a feature of the approximation and not a numerical problem.

\begin{figure}[htbp]
  \centering
  \includegraphics[width=1.0\columnwidth]{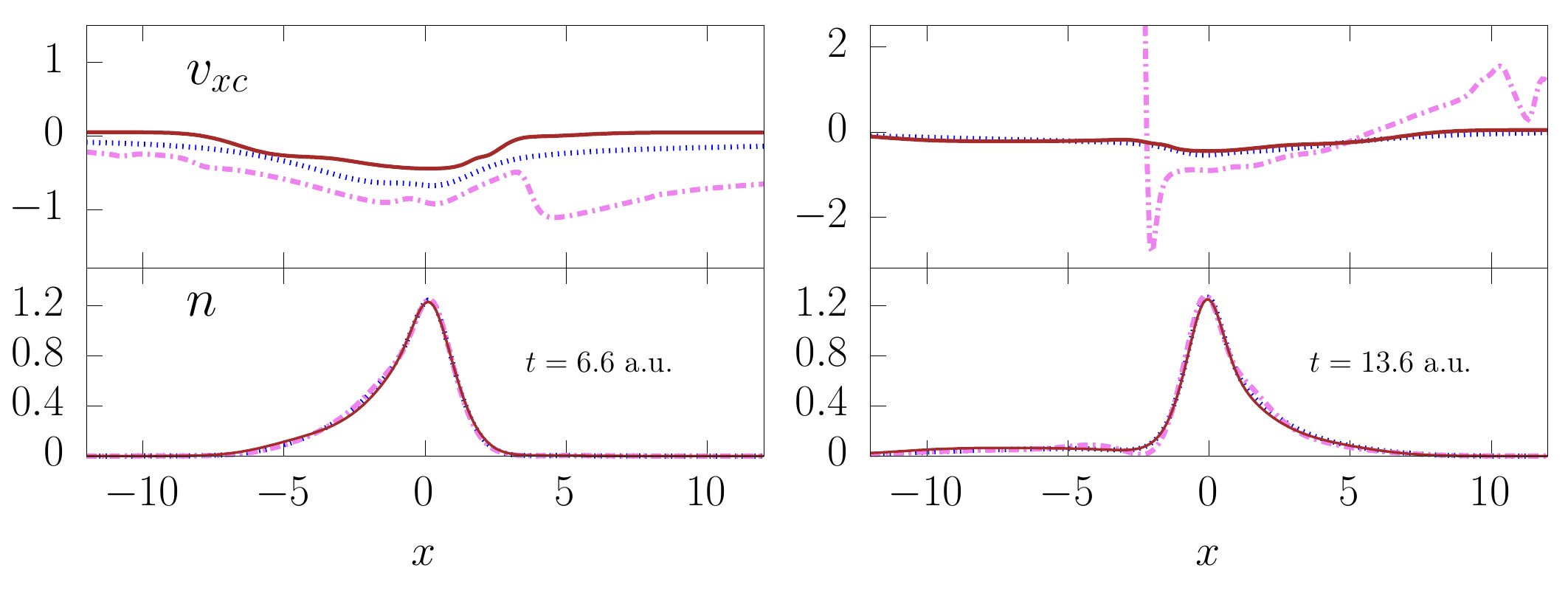}
  \caption{
    Time-snapshots of the xc potential (upper subpanels) and density (lower panels) for  
    propagation of $4$ electrons prepared in a LDA groundstate and evolved in a non resonant laser field: 
    $v\subxc\S$ approximation (blue dotted), $v\subxc^{\trho}$ (pink dot-dashed), and LDA (brown solid). 
  }
  \label{Fig:4e_laser_pot}
\end{figure}

As the diverging structures come from these sharp minima that develop in the density,
we tried a system with more electrons, hoping the density to be smoother in this case. 
To do so, we chose the external potential to be $v\subext(x) = -4/\sqrt{1+x^2}$,
and prepared the system with $4$ electrons in its LDA groundstate, i.e. two doubly-occupied orbitals.
The system is then excited with a non-resonant laser field $V(x,t) = 0.1sin(0.4t)x$.
As shown on Fig.~\ref{Fig:4e_laser_pot}, the initial potential develops the familiar step structure that appears when systems get driven far from equilibrium,
evident here at $t=6.6$ a.u.
Sadly,the hope is dashed that generally more electrons will buffer the development of the diverging structure: we observe at $t = 13.6$ a.u. once again the sharp needle structure has appeared, concomitant with a sharp minimum in the density. In fact in the other approximations shown, ALDA and $v\subxc\S$, there is not such a pronounced minimum; this minimum and the sharp needle structure appear to be self-consistent, self-amplifying, features.
In this dynamics, the troublesome region is occupied largely by the outer electrons; the two core electrons are only slightly perturbed by the field, and so do not act as an effective buffer.  
A movie is provided in the supplementary information.

Finally, we briefly discuss the case of electron-scattering off a 
 hydrogen atom initially in its ground state (fourth movie in the supplementary material).
The $v\xc^{\trho}$ propagation reproduces the density and exact potential accurately, even capturing the large peak behind the target atom on electron approach, until again the calculation is killed by a needle structure.


\section{Conclusions and outlook}

In conclusion, we have presented a new, first-principles approach to developing non-adiabatic functionals, that begins with an exact expression for the time-dependent xc potential, and, in contrast to previous approaches, does not simple bootstrap a ground-state approximation. The KS equation is coupled to an equation for a 1RDM, with information feeding back and forth between the equations, as the system is solved. 
 The dynamical step and peak features which have remained elusive, appear even when a simple HF-driven approximation is made in the 1RDM equation; these features arise from the 
  the form of the equation
    that defines $v\c^T$. The initial approximation explored here satisfies the ZFT, the GTI, is self-interaction-free, and contains memory, however after a short time unphysical instabilities appear. A better understanding of these divergences could lead to a better $\trho_2[\trho_1,\rho\1s]$, and work is underway in this direction. The general approach also opens the door to possibly by-passing the 1RDM calculation altogether, for example by searching for a paradigmatic system for which one can obtain the interacting 1RDM as a functional of KS accessible quantities.

Financial support from the US National Science Foundation
CHE-1566197 (L.L.) and the Department of Energy, Office
of Basic Energy Sciences, Division of Chemical Sciences,
Geosciences and Biosciences under Award DE-SC0015344 (N.T.M)
are gratefully acknowledged. 

\section{Supporting information}

{\bf movie\_a1\_rhotilde\_VxcS\_hf.mpg} :
Time evolution of the density (upper panel)  and xc potential (lower panel)  for field-free propagation of the state Eq.~(\ref{Eq:res_50:50}), predicted by KS evolution of $a=1$ KS state under the following approximations: exact $v\subxc$ (black), $v\subxc\S$ approximation (green), $v\subxc^{\trho}$ (pink), and HF-driven (orange solid).\\
{\bf movie\_a0\_rhotilde\_VxcS\_hf.mpg} :
Time evolution of the density (upper panel)  and xc potential (lower panel)  for field-free propagation of the state Eq.~(\ref{Eq:res_50:50}), predicted by KS evolution of $a=0$ KS state under the following approximations: 
exact $v\subxc$ (black), $v\subxc\S$ approximation (green), $v\subxc^{\trho}$ (pink), and HF-driven (orange solid).\\
{\bf movie\_4e\_laser\_rhotilde\_eps1d-3.gif} : 
Time evolution of the density (upper panel)  and xc potential (lower panel)  for  
propagation of $4$ electrons prepared in a LDA groundstate and evolved 
in a non resonant laser field: 
$v\subxc\S$ approximation (blue dotted), 
$v\subxc^{\trho}$ (pink dot-dashed), and LDA (brown solid).\\
{\bf movie\_scatt\_rhotilde\_VxcS.mpg} : 
Time evolution of the density (upper panel) and xc potential (lower panel) for
the scattering problem. 
The initial KS wavefunction is chosen identical to the interacting one:
$\Psi(x_1,x_2;t=0) = \Phi(x_1,x_2;t=0)$ where $\Phi(x_1,x_2;t=0)=   
  \frac{1}{ \sqrt{2}}
  \left(\phi_{\sss gs}(x_1)\phi_{\sss WP}(x_2) 
    +
\phi_{\sss WP}(x_1)\phi_{\sss gs}(x_2) \right) \quad $
and $\phi_{\sss gs}$ is the groundstate of the soft-Coulomb Hydrogen atom and 
$\phi_{\sss WP}$ is the incoming wavepacket defined as
$ \phi_{\sss WP}(x)=
  \left(0.2/\pi\right)^{\frac{1}{4}}e^{\left[-0.1(x-10)^2+i1.5(x-10)\right]}$. 
We display the exact propagation $v\subxc$ (black), the $v\subxc\S$ approximation (green) and the $v\subxc^{\trho}$ approximation (pink). 


\providecommand{\latin}[1]{#1}
\providecommand*\mcitethebibliography{\thebibliography}
\csname @ifundefined\endcsname{endmcitethebibliography}
  {\let\endmcitethebibliography\endthebibliography}{}

\end{document}